\documentclass{rmf-d}
\usepackage{nopageno,rmfbib,multicol,times,epsf,amsmath,amssymb,cite}
\usepackage[latin1]{inputenc}
\usepackage[]{caption2}
\usepackage{color}
\usepackage{graphicx}
\def\rmfcornisa{Research OR Education of Physics \hfill\rmf\ {\bf ??} (*?*) ???--???
\hfill MES? A\~NO?}

\def\rmfcintilla{{\it Rev.\ Mex.\ Fis.\/} {\bf ??} (*?*) (????) ???--???}
\clearpage \rmfcaptionstyle 
\begin{document}
\title{New Baryon States in Exclusive Meson Photo-/Electroproduction with CLAS 
\vspace{-6pt}}
\author{Victor I. Mokeev and Daniel S. Carman}
\address{12000 Jefferson Ave., Newport News, VA 23606, Jefferson Laboratory\\
(for the CLAS Collaboration)}
\maketitle
\recibido{day month year}{day month year \vspace{-12pt}}
\begin{abstract}
\vspace{1em} Impressive progress achieved in the past decade in experimental studies of exclusive meson photoproduction off protons 
and global multi-channel amplitude analyses has resulted in the discovery of several long-awaited new  nucleon resonances, with a 
decisive impact from the results of $K\Lambda$ and $K\Sigma$ photoproduction measured with the CLAS detector at Jefferson Lab. Further 
extension of these efforts towards combined studies of exclusive meson photo- and electroproduction data off protons will be presented. 
A new excited state of the nucleon, the $N'(1720)3/2^+$, discovered from combined analyses of $\pi^+\pi^-p$ photo- and electroproduction 
data, in addition to new resonances discovered in photo- and hadroproduction data, demonstrate the promising prospects of this new 
research avenue for discovery of additional new resonances. 
\vspace{1em}
\end{abstract}
\keys{ New baryon states, Exclusive meson photo- and electroproduction, Charged double pion photo- and electroproduction \vspace{-4pt}}
\pacs{\bf{\textit{75.25.-j, 13.60.-r, 13.88.+e, 24.85.+p}} \vspace{-4pt}}
\begin{multicols}{2}

\section{Introduction}

Studies of the spectrum of excited nucleon states ($N^*$) shed light on approximate symmetries of the strong interaction in the regime of 
large (comparable with unity) QCD running coupling, the so-called strong QCD regime, which underlies the generation of the $N^*$ spectrum 
\cite{sqcd19,mokeev2020,capst2000,cap-is1986,sant2015}. The full $N^*$ spectrum of nature, including those states already observed and those
that are still to be discovered, defines the rate for the transition from the primordial deconfined mixture of quarks and gluons into the 
hadron gas phase that took place within the first microseconds after the Big Bang~\cite{baz2014,baz2014a,bur2020}. In this phase transition 
the dominant part of hadron mass was generated, chiral symmetry of QCD was broken dynamically, and quark-gluon confinement emerged. These 
features define the essence of the strong QCD regime that makes the studies of the $N^*$ spectrum a compelling experimental program to 
explore the emergence of hadron matter from QCD. The recent advances in the search for new excited states of the nucleon, also known as the 
``missing" resonances, will be presented in this proceedings. 

\section{``Missing" Resonances from Exclusive Meson Photoproduction Data}

\begin{figure*}
\centering
\includegraphics[width=0.75\textwidth]{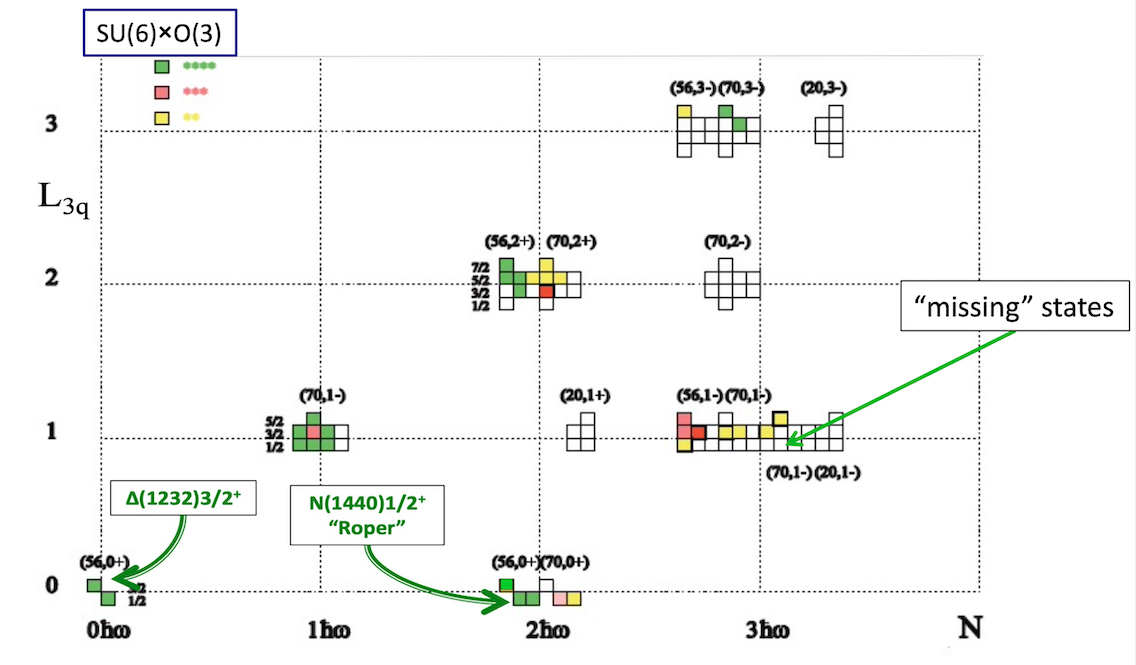}
\vspace{-2mm}
\caption{Spectrum of nucleon resonances expected in quark models employing ${\rm SU(6)}_{\rm spin-flavor} \times {\rm O(3)}_{\rm space}$ 
symmetry. $L_{3q}$ is the orbital angular momentum of the three constituent quarks and the quantum number $N$ corresponds to the radial 
excitation of the three-quark system. The predicted and observed states are shown by the filled boxes, while the predicted and still not 
observed states are shown by the open boxes.}
\label{pred_spectr}
\end{figure*}

Constituent quark models based on approximate symmetries of the strong interaction that are relevant for the strong QCD regime and established 
by analyzing the $N^*$ spectrum known before 2012~\cite{capst2000,cap-is1986,karl-is1982,sant2015,klempt2010}, predict many more excited states 
of the nucleon than have been observed in experiments both with electromagnetic and hadronic probes. The expectation from quark models that 
employ SU(6)$\times$O(3) (spin-flavor$\times$space-rotational) symmetry is depicted in Fig.~\ref{pred_spectr}. The predicted and observed 
nucleon resonances are shown by the filled boxes. The states that are predicted but still not observed are shown by the open boxes. The search 
for the many states in the mass range above 1.7~GeV that have eluded detection has become the focus of the extensive studies to address the 
so-called ``missing" resonance problem. Quark model expectations of the $N^*$ spectrum starting from the QCD Lagrangian both within lattice and
continuum QCD approaches support the states predicted from SU(6)$\times$O(3) symmetry expectations~\cite{edw2011,chen2019,qin2019}. The studies 
of exclusive meson photoproduction extend the capabilities to search for these resonances in comparison with the results available from exclusive 
meson hadroproduction with the biggest contribution from data with pion beams. Exclusive meson production with pion beams is sensitive to the
resonances with substantial decay into the $N\pi$ final states, while exclusive photoproduction processes allow us to pin down the resonances 
with decays into $N\pi$~\cite{arn-07} as well as other final hadron states such as $K\Lambda$, $K\Sigma$, and $N\pi\pi$
\cite{brad2007,mcc2009,dey2010,gol2019}. According to the quark models~\cite{capst2000,sant2015,bijk2016}, the non-$N\pi$ final states can 
strongly couple to the ``missing" resonances. The search for these new states has driven the exploration of the $N^*$ spectrum in experiments 
with electromagnetic probes for the past two decades~\cite{bu2004}.

\begin{figure*}
\centering
\includegraphics[width=0.75\textwidth]{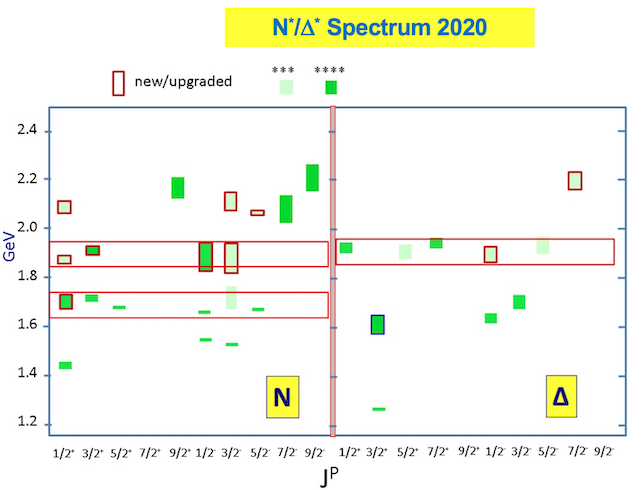}
\caption{The $N^*$ spectrum established in global multi-channel analyses of exclusive meson photo- and hadroproduction data~\cite{bur2020}. 
The recently discovered new resonances are highlighted with the brown boxes.}
\label{mes_spectr}
\end{figure*}

Recently, several long-awaited new nucleon resonances were discovered in global multi-channel analyses of exclusive meson photo- and
hadroproduction data~\cite{anis2017,ron2014} with a decisive impact of the CLAS results on $K\Lambda$ and $K\Sigma$ photoproduction
\cite{brad2007,mcc2009,dey2010}. Implementation of new nucleon resonances in the mass range from 1.8~GeV to 2.2~GeV is essential to 
describe the data on the $K\Lambda$ and $K\Sigma$ differential cross sections and induced asymmetries at backward angles from CLAS. After 
the implementation of the new resonances, a good description of the observables for most exclusive photo- and hadroproduction channels 
relevant in the resonance region and included into the coupled-channel approaches~\cite{anis2017,ron2014} was achieved, providing strong 
evidence for the existence of these new states. The established $N^*$ spectrum is shown in Fig.~\ref{mes_spectr} with the recently 
discovered states highlighted with the brown boxes. Two recently discovered resonances, the $N(1895)1/2^-$ and $N(1900)3/2^+$, have been
assigned the highest four-star PDG status~\cite{pdg} as firmly established resonances. Knowledge on other recently observed nucleon 
resonances has been greatly improved as reflected by their current PDG status (increased from two stars to three). The discovery of these 
new long-awaited resonances represents an important achievement in hadron physics. 

\section{The $N^*$ Spectrum from Combined Studies of Exclusive Meson Photo- and Electroproduction Data}

Combined studies of exclusive meson photo- and electroproduction data open up new prospects in the exploration of the $N^*$ spectrum. New 
nucleon resonances seen in photoproduction can be also observed in electroproduction. The resonance masses and the total/partial hadronic 
decay widths obtained in analyses of exclusive electroproduction data should be the same as established from the exclusive photoproduction 
data. A successful description of the exclusive meson photo- and electroproduction data within a broad range of photon virtualities $Q^2$ 
with $Q^2$-independent nucleon resonance masses, and total and partial hadronic decay widths, will validate the resonance existence in a 
nearly model-independent way. The new $N'(1720)3/2^+$ resonance was recently discovered in the combined studies of the CLAS $\pi^+\pi^-p$ 
photo- and electroproduction data~\cite{mok2020} in addition to new resonances established in the analysis of exclusive meson photo- and 
hadroproduction data~\cite{anis2017}.

\begin{figure*}
\centering
\includegraphics[width=0.8\textwidth]{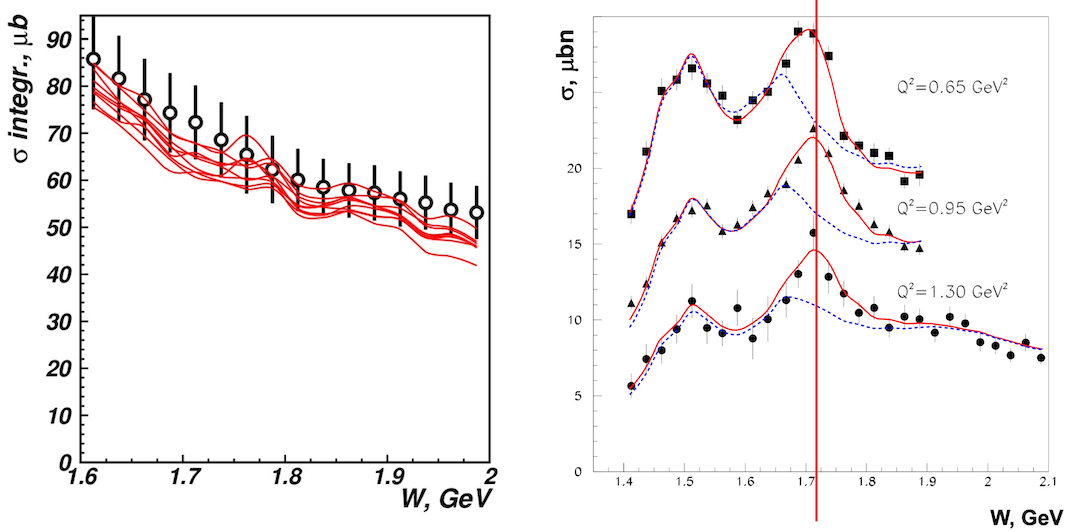}
\vspace{-2mm}
\caption{Fully integrated $\pi^+\pi^-p$ photo- (left) and electroproduction (right) cross sections as a function of $W$. The photoproduction 
data points~\cite{gol2019} are shown by the open circles with the statistical and systematic uncertainties added in quadrature. The 
electroproduction data points are shown by the filled squares with only statistical uncertainties represented. The series of red curves in 
the left panel represents the computed cross sections within the JM model~\cite{mok2008,mok2012,mok2016} selected in the data fit including 
the new $N'(1720)3/2^+$. The resonant/non-resonant parameters of the JM model are fit to the data on nine one-fold differential cross sections 
in each bin of $(W,Q^2)$ covered by the measurements~\cite{rip2003,gol2019}. The description of the fully integrated $\pi^+\pi^-p$ 
electroproduction cross sections within the JM model accounting for the $N'(1720)3/2^+$ is shown by the red lines in the right panel. The 
blue dashed curves represents the JM model results from the 2003 version~\cite{mok2003,rip2000} accounting for only conventional resonances 
with the $N(1720)3/2^+$ partial decay width into $\rho p$ from the 2002 PDG listings.}
\label{npipi_electro_photo}
\end{figure*}

\begin{figure*}
\centering
\includegraphics[width=0.8\textwidth]{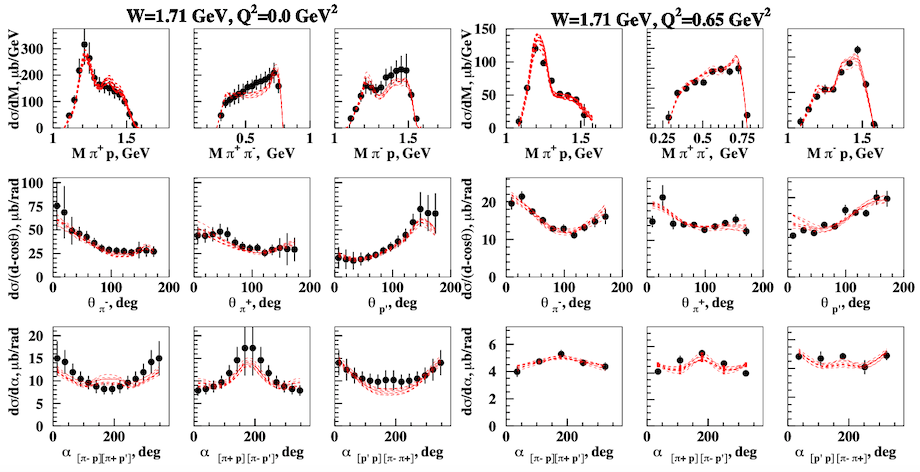}
\vspace{-2mm}
\caption{Representative examples of the description of the $\pi^+\pi^-p$ one-fold differential cross sections achieved within the JM model
\cite{mok2008,mok2012,mok2016} after implementation of the new $N'(1720)3/2^+$. The computed cross sections selected in the data fit are 
shown by the red curves.}
\label{one_diff}
\end{figure*}

Resonance-like structures were observed a long time ago in the $W$-dependence of the fully integrated $\pi^+\pi^-p$ electroproduction cross 
sections from CLAS~\cite{rip2003} in the third resonance region (see Fig.~\ref{npipi_electro_photo} with the peak positions at 
$W \approx 1.71$~GeV in all three $Q^2$-bins of this measurement). Recently, data on the $\pi^+\pi^-p$ photoproduction cross sections were 
obtained for $W$ from 1.6-2.0~GeV~\cite{gol2019}. In order to explore the resonance contributions into the $\pi^+\pi^-p$ photo- and 
electroproduction cross sections in the third resonance region, we analyzed nine independent one-fold differential photo-/electroproduction 
cross sections (see representative examples in Fig.~\ref{one_diff}). The analysis was carried out within the $W$-interval from 1.60~GeV to 
1.76~GeV for $Q^2 < 1.5$~GeV$^2$ within the Jefferson Lab/Moscow State University (JM) reaction model~\cite{mok2008,mok2012,mok2016} developed 
for the extraction of the $\gamma_{r,v}pN^*$ photo-/electrocouplings from combined studies of the $\pi^+\pi^-p$ data. The JM model provides a 
good description of all data for $W$ up to 2.0~GeV and $Q^2$ from the photon point up to 5.0~GeV$^2$. All $\gamma_vpN^*$ electrocouplings 
obtained from the charged double pion electroproduction channel and included in the PDG have become available from the data analyses within 
the JM model.

Analysis of the data on the $\pi^+$, $\pi^-$ and $p$ center-of-mass (CM) angular distributions reveals essential contributions from resonances 
of spin-parity $J^P=3/2^+$ to $\pi^+\pi^-p$ photo- and electroproduction in the third resonance region. Therefore, we explored two possibilities 
to describe the $\pi^+\pi^-p$ data in this region: a) either accounting for only conventional resonances with a substantial contribution from 
the conventional $N(1720)3/2^+$ or b) by implementing on top of the conventional resonances a contribution from a new resonance labeled as 
$N'(1720)3/2^+$ with mass, $\pi\Delta$ and $\rho p$ hadronic decay widths, and photo-/electrocouplings determined from a combined fit of the
$\pi^+\pi^-p$ photo- and electroproduction cross sections. In the data fit, we simultaneously varied the parameters of the nucleon resonances 
and the parameters of the non-resonant mechanisms included in the JM model. Eventually, we selected the computed cross sections with minimal 
values of $\chi^2/$(data point) in relation to the data. The selected computed cross sections were spread within the data uncertainties for 
most data points covered by the measurements. Under both assumptions (a) and (b) on the resonant contributions, a good description of the data
was achieved. Representative examples are shown in Figs.~\ref{npipi_electro_photo} and ~\ref{one_diff}. The values of $\chi^2$/(data~point) 
obtained from comparison between the measured and computed nine one-fold differential cross sections are comparable for both assumptions.  

For the case when the contributions from only conventional resonances are taken into account, the branching fraction for the decay of the
conventional $N(1720)3/2^+$ resonance into the $\rho p$ final state inferred from the fits of the $\pi^+\pi^-p$ photo- and electroproduction 
data are different by more than factor of 4 (see Table~\ref{conventional}). This contradiction conclusively demonstrates that accounting 
only for the contributions from conventional resonances does not allow us to consistently describe both the $\pi^+\pi^-p$ photo- and 
electroproduction data. Implementation of the new $N'(1720)3/2^+$ state makes it possible to describe the $\pi^+\pi^-p$ photo- and
electroproduction data over the broad range of $Q^2$ from 0 to 1.5~GeV$^2$. A good data description was achieved with $Q^2$-independent 
masses, with consistent $\pi \Delta$ and $\rho p$ hadronic decay widths of the new $N'(1720)3/2^+$ state and of the other conventional 
resonances with substantial contributions into the $\pi^+\pi^-p$ channel in the third resonance region (see Table~\ref{conv_miss}). This 
provides strong evidence for the existence of the $N'(1720)3/2^+$ resonance.

\begin{figure*}
\centering
\includegraphics[width=0.8\textwidth]{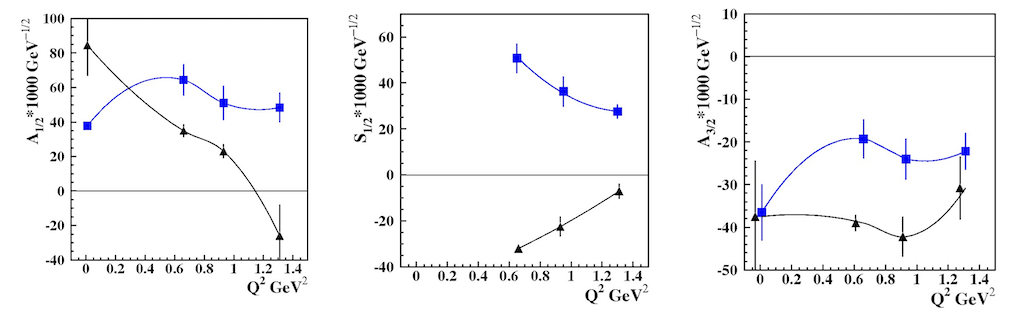}
\vspace{-2mm}
\caption{The $\gamma_{r,v}pN^*$ photo-/electrocouplings of the conventional $N(1720)3/2^+$ (black) and the new $N'(1720)3/2^+$ (blue) 
resonances from the $\pi^+\pi^-p$ photo-~\cite{gol2019} and electroproduction~\cite{rip2003} data fits.}
\label{conv_miss_electrocoupl}
\end{figure*}

\begin{table*}[htb!]
\begin{center}
\begin{tabular}{|c|c|c|c|} \hline
 {$N$(1720)3/2$^+$} & $N^*$ total width       & Branching fraction            & Branching fraction         \\
               & MeV                     &  for decays to $\pi\Delta$    &  for decays to $\rho N$    \\ \hline
 electroproduction    &  126.0 $\pm$ 14.0       &    64\% - 100\%               & $<$5\%                    \\
 photoproduction      &  160.0 $\pm$ 65.0       &   14\% - 60\%                 &  19\% - 69\%               \\ \hline
\end{tabular}
\end{center}
\caption{$N$(1720)3/2$^+$ hadronic decay widths and branching fractions into $\pi \Delta$ and $\rho p$ determined from independent fits to 
the data on charged double-pion photo-~\cite{gol2019} and electroproduction~\cite{rip2003} off protons accounting only for contributions 
from previously known resonances.}
\label{conventional}
\end{table*}

\begin{table*}[htb]
\begin{center}
\begin{tabular}{|c|c|c|c|} \hline
Resonance            & $N^*$ total width         & Branching fraction            & Branching fraction         \\
states                & MeV                       & for decays to $\pi\Delta$     &  for decays to $\rho p$    \\ \hline
 $\Delta(1700)3/2^-$  &                           &                               &                           \\
 electroproduction    &  288.0 $\pm$ 14.0         &    77 - 95\%                    &   3 - 5\%                      \\
 photoproduction      &  298.0 $\pm$ 12.0         &    78 - 93\%                    &   3 - 6\%                    \\ \hline
 N(1720)3/2$^+$       &                           &                               &                           \\
 electroproduction    &  116.0 $\pm$ 7.0          &    39 - 55\%                    &   23 - 49\%                     \\
 photoproduction      &  112.0 $\pm$ 8.0          &    38 - 53\%                    &   31 - 46\%                    \\ \hline
N$^{\, '}$(1720)3/2$^+$ &                          &                               &                           \\
 electroproduction    &  119.0 $\pm$ 6.0         &     47 - 64\%                  &    3 - 10\%                     \\
 photoproduction      &  120.0 $\pm$ 6.0         &     46 - 62\%                   &    4 - 13\%                    \\ \hline
\end{tabular}
\end{center}
\caption{Hadronic decays into the $\pi \Delta$ and $\rho p$ final states of the resonances in the third resonance region with major decays
to the $\pi^+\pi^- p$ final state determined from the fits to the data on charged double-pion photo-~\cite{gol2019} and electroproduction
\cite{rip2003} after implementing a new $N'(1720)3/2^+$ baryon state.}
\label{conv_miss}
\end{table*}

The masses, total decay widths, and branching fractions for hadronic decays into the $\pi\Delta$ and $\rho p$ final states for the 
conventional $N(1720)3/2^+$ and for the new $N'(1720)3/2^+$ resonances inferred from the data fit are presented in Table~\ref{miss_conv_hadr}.
The $\gamma_{r,v}pN^*$ photo-/electrocouplings of these states derived from fitting the nine independent one-fold $\pi^+\pi^-p$
photo-/electroproduction cross sections are shown in Fig.~\ref{conv_miss_electrocoupl}.

\begin{table*}
\begin{center}
\begin{tabular}{|c|c|c|c|c|} \hline
Resonance        & Mass,       & $N^*$ total width, & Branching fraction            & Branching fraction         \\
states              & GeV          & MeV                        &  for decays to $\pi\Delta$ &  for decays to $\rho p$ \\
\hline
$ N(1720)3/2^+$ & 1.743-1.753 &       114 $\pm$ 6       & 38-53\%                        &   31-46\%                     \\
\hline
$N'(1720)3/2^+$ & 1.715-1.735 &       120 $\pm$  6      & 47-62\%                         &    4-10\%                     \\
\hline
\end{tabular}
\end{center}
\caption{Masses and hadronic decay widths of the $N(1720)3/2^+$ and $N'(1720)3/2^+$ resonances to the $\pi \Delta$ and $\rho p$ final states
determined as the overlap between the parameter ranges from independent fits of the $\pi^+\pi^-p$ photo- and electroproduction data
\cite{rip2003,gol2019}.}
\label{miss_conv_hadr}
\end{table*}

We found that for the successful description of the combined $\pi^+\pi^-p$ data, the contributions from both the conventional $N(1720)3/2^+$ 
and the new $N'(1720)3/2^+$ resonances are needed. These two excited states of the nucleon have almost the same mass, and their total
decay widths overlap within the uncertainties for their total hadronic decay widths. However, they have different branching fractions for 
decays into $\pi\Delta$ and $\rho p$ (see Table~\ref{miss_conv_hadr}). They also have different dependencies of their electrocouplings on $Q^2$ 
(see Fig.~\ref{conv_miss_electrocoupl}). The differences in their hadronic decays prevent mixing between the $N'(1720)3/2^+$ and $N(1720)3/2^+$ 
states of the same isospin, spin, and parity. Evidence for two resonances of spin parity $J^P=3/2^+$ and of isospin 1/2 was obtained in global
coupled-channel analyses of exclusive photo-/hadroproduction data~\cite{kamano2013}. Furthermore, it was found that the resonance-like 
structures observed in the $W$-dependence of the inclusive electron scattering cross section in the third resonance region are created with 
the biggest contribution from the new $N'(1720)3/2^+$ state~\cite{blin2021}, which supports the existence of this new resonance. 

Currently the $N'(1720)3/2^+$ is the only new baryon state for which the results on the $Q^2$ evolution of the $\gamma_vpN^*$ electrocouplings 
are available, offering insight into the internal structure of previous ``missing" resonances and allowing us to shed light on the peculiar 
features in their structure that have made their observation so elusive. The electrocouplings of both the $N(1720)3/2^+$ and $N'(1720)3/2^+$
computed within Ads/CFT~\cite{lub2020} represent a first promising step in this direction. Analysis of the high-lying nucleon resonance 
spectrum, as we know it in 2020, suggests SU(6) spin-flavor assignments for the conventional $N(1720)3/2^+$ and the new $N'(1720)3/2^+$ 
resonances as [56,2$^+$] and [70,2$^+$], respectively. These assignments imply: a) both resonances consist of three constituent quarks with 
orbital angular momentum $L$=2 and of total quark spin $S_q$=1/2~\cite{mok2020}, b) the new $N'(1720)3/2^+$ resonance represents a system of
three bound quarks with orbital excitation over both the $\rho$ and $\lambda$ coordinates for the three-body system that has been observed
for the first time. New data on the $N(1720)3/2^+$ and $N'(1720)3/2^+$ electrocouplings at $Q^2$ up to 5.0~GeV$^2$ foreseen from CLAS 
analyzed within different quark models will allow us to pin down peculiar features in the structure of new baryon states.

\section{Conclusions and Outlook}

Several long-awaited new ``missing" nucleon resonances have been discovered from global analyses of exclusive meson photo- and hadroproduction 
data with a decisive impact from the $K\Lambda$ and $K\Sigma$ photoproduction channels measured with CLAS. A new $N'(1720)3/2^+$ resonance has 
been observed in the combined studies of $\pi^+\pi^-p$ photo- and electroproduction data, and this state is the only ``missing" resonance for
which the results on the $Q^2$-evolution of the $\gamma_vpN^*$ electrocouplings have become available. In the future, the information on the
$N'(1720)3/2^+$ electrocouplings from the CLAS data will be extended towards $Q^2$ up to 5.0~GeV$^2$. Analyses of the results on the new 
resonance electrocouplings in collaborative efforts with hadron structure theory will shed light on the particular features of the ``missing" 
resonance structure that have made them so elusive for detection. 

\section{Acknowledgments}

This material is based upon work supported by the U.S. Department of Energy, Office of Science, Office of Nuclear Physics under contract
DE-AC05-06OR23177. The U.S. Government retains a non-exclusive, paid-up, irrevocable, world-wide license to publish or reproduce this 
manuscript for U.S. Government purposes. 


\end{multicols}
\medline
\begin{multicols}{2}

\end{multicols}
\end{document}